\documentclass[aip,reprint]{revtex4-1}

\usepackage{graphicx}
\usepackage{dcolumn}
\usepackage{bm}
\usepackage{amsmath}

\usepackage[utf8]{inputenc}
\usepackage[T1]{fontenc}
\usepackage{mathptmx}
\usepackage{etoolbox}
\usepackage{siunitx}
\usepackage[caption=false]{subfig}
\draft 

\makeatletter
\def\@email#1#2{%
 \endgroup
 \patchcmd{\titleblock@produce}
  {\frontmatter@RRAPformat}
  {\frontmatter@RRAPformat{\produce@RRAP{*#1\href{mailto:#2}{#2}}}\frontmatter@RRAPformat}
  {}{}
}%
\makeatother

\begin{document}

\preprint{Preprint - Technical University of Darmstadt, Institute for Accelerator Science and EM Fields}

\title{Simulation of trapped non-neutral plasma dynamics with rotating wall compression} 

\author{L. Riik}
\affiliation{Technical University of Darmstadt, Institute for Accelerator Science and EM Fields,Schloßgartenstraße 8, 64289 Darmstadt, Germany}
\author{O. Boine-Frankenheim}
\affiliation{Technical University of Darmstadt, Institute for Accelerator Science and EM Fields,Schloßgartenstraße 8, 64289 Darmstadt, Germany}
\affiliation{ GSI Helmholtzzentrum für Schwerionenforschung GmbH, Planckstraße 1, 64291 Darmstadt, Germany }

\email{luisa.riik@temf.tu-darmstadt.de}

\date{\today}

\begin{abstract}
A spectral macroparticle simulation scheme is used to model non-neutral plasma dynamics in a Penning-Malmberg trap. This grid-free numerical scheme is well adapted to the specific geometry of trapped plasmas, with varying longitudinal and transverse profiles in a cylindrical conducting pipe. The cutoff spectral harmonics provide control over the trade-off between accuracy and noise smoothing, which is especially important in the three-dimensional case. We demonstrate the ability of this scheme to obtain the dispersion relation of the plasma modes from the inherent simulation noise, providing a valuable tool for understanding the plasma's behavior. In the presence of a rotating wall drive, we show that the numerical scheme can reproduce the time evolution of the driven eigenmodes, identify the limit of the linear response with increasing drive amplitude and retrieve the compression rate dependency on the drive frequency, known from the weak drive regime of the rotating wall technique. Since the model contains no cooling, drive-induced heating restricts the simulations to short times and the weak-drive regime. To the authors’ knowledge, such simulations have not been reported in the literature so far.
\end{abstract}


\maketitle 

\section{\label{sec: Introduction}Introduction}

    A Penning Malmberg trap is a device widely used to confine charged particles for a variety of physical applications. In the simplest case, it is composed of three conducting cylinders where the left and right one are set on an electric potential which confines the particles in axial direction. Additionally, there is a static homogeneous magnetic field aligned to the axial direction, which confines the particles in radial direction due to Lorentz forces. Especially for investigations of and experiments with antimatter, this became an important tool for accumulation and storage. One of these experiments is the PUMA (antiProton Unstable Matter Annihilation) experiment \cite{Aumann2019}, which aims to investigate the proton to neutron density ratio of exotic nuclei by means of low-energy antiproton annihilation process in the tail of the nuclear density. To make this possible, the antiprotons have to be transported to the facility where the exotic nuclei are produced. The transport and the time needed for accumulation and experiments require very long confinement times of the antiprotons. 
    However, long confinement times are impeded by particle loss due to absorption by the trap wall. Collisions and imperfections in the experimental setup, like field asymmetries, will cause the particle cloud to expand, so that the particles will eventually reach the wall of the trap. To act against these expansion mechanisms, a rotating electrostatic field can be employed. This technique is called rotating wall drive and is used routinely in many applications to realize very long-confinement times, e.g.~\cite{Huang1997,Anderegg1998,Hollmann2000}. It is realized by an azimuthally segmented electrode. A sinusoidal voltage is applied with appropriate phase differences between the segments, and depending on these phase differences a rotating electrostatic dipole field or other multipole fields are created.  

    Rotating-wall compression is a standard tool throughout the non-neutral plasma community. In the antiproton and antihydrogen experiments at the CERN Antiproton Decelerator, rotating fields are routinely used to compress and stabilize electron, positron and antiproton plasmas before and during antihydrogen formation \cite{ALPHA2008,ALPHA2018,Farjans2020}. They are equally central to the accumulation of positrons and to trap-based positron beams for atomic, molecular and materials science \cite{Greaves2000, Danielson2015}, and to the preparation of well-controlled plasmas for precision measurements and for fundamental studies of plasma physics \cite{Maero2024}. In all these applications, the achievable density and the heating during compression depend on how the rotating field couples to the plasma modes, which is difficult to access experimentally. A mode-resolved kinetic description of this coupling is therefore of interest beyond a single experiment.
    
    There has been theoretical work on the compression by rotating fields \cite{Eggleston1999,Kiwamoto2005, Lazarow_2023} where it is shown that this effect can be described in the four dimensional phase space, all three space dimensions, and the axial velocity. This is the lowest dimensionality to which one can reduce a simulation model. The rotating wall drive can be divided into two different regimes. The weak drive for low drive amplitudes couples to the Trivelpiece-Gould (TG) modes. These excited modes in turn, exert a torque on the plasma column by Landau resonance mechanisms \cite{Kiwamoto2005}. Therefore, in this regime there is only compression if the drive frequency is close to a TG-mode. On the other hand, for higher drive amplitudes, in the strong drive regime, there is no dependencies on the drive frequency observable. Understanding the response of the trapped plasma to external fields is crucial for its control and manipulation. Density, temperature and shape of the trapped plasma, as well as the trap geometry, highly influence how the trapped plasma reacts to an external excitation. 

    Although there are plenty of experimental investigations of this rotating wall technique, there are no simulations based on theoretical models, known to the author, that describe the process from coupling of the rotating fields to plasma waves and subsequent compression or expansion. Theoretical models often rely on simplifying assumptions, while experimental measurements are limited by the difficulty of accessing information on specific plasma parameters. In this context, numerical simulations offer a powerful tool for exploring the behavior of trapped plasmas in a controlled and systematic manner. 
    
    In this publication, we use a spectral macro-particle simulation scheme for studying the behavior of trapped plasmas, with a focus on the effects of external excitations on the plasma's eigenmodes. A Fourier-Bessel decomposition can be used to calculate the space charge field of macro-particles in a cylindrical tube, as was demonstrated for intense charged particle beams in a conducting pipe\cite{Qiang2001,Batygin2001}, a system that is analogous to non-neutral plasma in a Penning-Malmberg trap. The longitudinal and angular decomposition of the space charge solver coincide with the spatial eigenmodes of the plasma model in this work. Additionally, inside the plasma column the radial dependency of the eigenmodes can be described by a series of Bessel functions.
    Hence, the longitudinal and azimuthal eigenmode structure can be read out directly from the space-charge solver, while the radial eigenmode amplitudes follow from a projection onto the radial eigenfunctions (Appendix A). This, in turn, provides the means to track the eigenmodes directly and investigate their response to an external excitation. The code is publicly available \cite{simulation_code}. For more details on the simulation scheme, see appendix \ref{sec_simulation_model}.
    
    The contributions of the present work are (i) the application of a grid-free drift-kinetic macro-particle scheme to Penning–Malmberg traps (ii) the extension of the linear response of driven eigenmodes \cite{wave_launching} to azimuthally non-symmetric drives; and (iii) its application to a mode-resolved study of rotating-wall compression in the weak-drive regime.

    This paper is organized as follows: First, the model of the trapped non-neutral plasma is explained in Sec. \ref{sec_nnp}, followed by a summary of the linear theory of TG-modes and an expression for the excitation of plasma eigenmodes by a rotating wall drive for the azimuthal nonsymmetric case in Sec. \ref{sec_lin_theory}. In Sec. \ref{sec_noise_spectra}, the spectral macroparticle simulation scheme is then applied to obtain the dispersion relation of the plasma eigenmodes from the inherent simulation noise. Thereafter, in Sec. \ref{sec_RW_response}, the response of the plasma eigenmodes to the rotating wall drive is studied, including the comparison to the expression for the rotating wall excitation and the dependence of the compression rate on the rotating wall drive frequency. In the present work, we restrict ourselves to the weak-drive regime and to times of a few plasma rotation periods. Since the model does not include cooling mechanisms, the heating by the drive limits both the duration and the amplitude of the simulated drive (see Sec. V). The strong-drive regime is beyond the scope of the present model.

\section{Non-neutral plasma and trap model \label{sec_nnp}}  
\begin{figure}
        \includegraphics[]{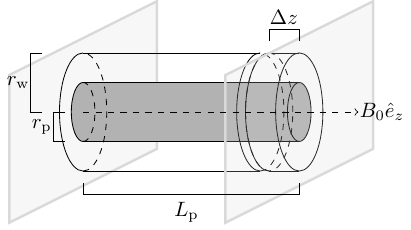}
        \caption{Schematic view of the model trap system. A plasma column (dark gray) with radius $r_{\mathrm{p}}$ inside a perfect conducting beam pipe with radius $r_{\mathrm{w}}$. The magnetic field is along the cylinder axis. The light gray areas indicate the periodicity length for an infinite column or the mirror ends in case of a finite column. For the rotating wall drive a part of the pipe with length $\Delta z$ is set to potential $V_{\mathrm{RW}}(\theta ,t)$.} 
        \label{fig: trap_model}
    \end{figure}

    The Penning-Malmberg trap is modeled as a perfect conducting, grounded cylindrical wall with radius $r_{\mathrm{w}}$, and a confining magnetic field aligned along the longitudinal axis $\mathbf{B}=B_0 \hat{e}_z$, as sketched in Fig.~\ref{fig: trap_model}.

    If the dynamics of the trapped particles are dominated by their space charge field, it is justified to describe them as a non-neutral plasma. The magnetic field is along the axis of the cylindrical electrodes. Transversal to the magnetic field, the particles perform cyclotron motions around the field lines with the modified cyclotron frequency $\omega_{\mathrm{c}} - \omega_{\mathrm{D}}$, which differs from the cyclotron frequency $\omega_{\mathrm{c}}=qB_0/m$ only by the small $E \times B$ rotation frequency $\omega_{\mathrm{D}} \ll\omega_{\mathrm{c}}$ \cite{Davidson}. Additionally, there is a $E \times B$ drift due to the transversal components of the space charge field. The radial component causes the drift centers to rotate around the cylinder axis with the plasma rotation frequency $\omega_{\mathrm{D}}$. Along the magnetic field lines, the particles bounce back and forth between the end electrodes with bounce frequency $\omega_{\mathrm{b}}$. In case of uniform charge distribution $\rho = q n_0$, the $E \times B$ or rotation frequency is given by $\omega_{\mathrm{D}} = q n_0 / 2\epsilon_0 B_0$.

    The plasma is assumed to be cold. This assumption results in the frequency scaling $\omega_{\mathrm{D}} \ll \omega_{\mathrm{p}} \ll \omega_{\mathrm{c}}$ and a very small ratio of the Debye length to other system lengths. These assumptions, found in many experimental cases, allow to use the drift kinetic approximation, where only the drift centers of the trapped particles are considered. In this approximation the collisionless drift kinetic Vlasov equation is\cite{Davidson}:
    \begin{equation}
       \left( \frac{\partial}{ \partial t}  + \frac{\mathbf{E} \times \mathbf{B}}{B_0^2} \nabla_{\perp} + v_z \frac{\partial}{ \partial z}  + \frac{q}{m_\mathrm{q}} \mathbf{E}_{\parallel} \frac{\partial}{\partial v_z} \right) f =0
       \label{equ: Vlasov_equation}
    \end{equation}
    where the distribution function $f(\mathbf{r},v_z,t)$ depends on time, spatial coordinates, and longitudinal velocity. Furthermore, the system can be considered to be non-relativistic and Poisson's equation is sufficient to calculate the self-field of the plasma:
    \begin{equation}
        \Delta\phi_{\mathrm{s.c.}}(\mathbf{r})= -\frac{\rho(\mathbf{r})}{\varepsilon_0}
    \end{equation}
    where $\rho(\mathbf{r})$ is the charge distribution of the drift centers.
    
    In equilibrium, the plasma column (dark gray in Fig.~\ref{fig: trap_model}) has a flat-top profile with constant density $n_0$ inside the radius $r_\mathrm{p}$ and a sharp edge. Additionally, the equilibrium plasma radius $r_{\mathrm{p}}$ and the trap radius $r_{\mathrm{w}}$ are considered small in relation to its length $L_{\mathrm{p}}$. Therefore, finite length effects will be neglected and the plasma can be modeled either as an infinitely long column with periodicity length $\mathcal{L}$ or as a finite column of length $L_\mathrm{p}$ with mirror boundary conditions. The light gray planes in Fig.~\ref{fig: trap_model} mark the boundaries of the periodicity length or the mirror planes, respectively. For the infinitely long plasma column case, Dirichlet boundary conditions at the trap wall and periodic boundary conditions in longitudinal direction are assumed for the self-field. The particles will be absorbed at the trap wall while being periodic in the longitudinal direction.
    The finite plasma column with specular ends can be constructed from the infinite long case. For that purpose, the plasma is mirrored around the transversal plane at $z=L_{\mathrm{p}}$ and the periodicity length is set to $\mathcal{L} = 2L_{\mathrm{p}}$ \cite{Eggleston1999,wave_launching}.

    The rotating wall drive is modeled by a ring electrode of length $\Delta z$, placed at one end of the finite plasma column (Fig.~\ref{fig: trap_model}).
    The voltage on the rotating wall electrode is described by
    \begin{equation*}
        V_{\mathrm{RW}}(\theta,t)=\phi_{\mathrm{rw}}(t)\cos{(m_{\mathrm{rw}} \theta - \omega_{\mathrm{rw}} t)}
    \end{equation*}
    with angular mode $m_{\mathrm{rw}} \geq1$ and frequency $\omega_{\mathrm{rw}}$. The resulting potential can be calculated analytically and is given by
    \begin{equation}
        \begin{split}
            \Phi_{\mathrm{RW}} = \sum_{n=1}^{\infty} \frac{I_{m_{\mathrm{rw}}} (k_n r)}{I_{m_{\mathrm{rw}}} (k_n r_{\mathrm{w}})} \frac{2(-1)^n}{\pi n}\phi_{\mathrm{rw}}(t)  \sin{(k_n \Delta z)}\\ \times \cos{(k_nz)} \cos{(m_{\mathrm{rw}}\theta - \omega_{\mathrm{rw}} t)} ~~,
        \end{split}
        \label{equ: RW-field}
    \end{equation}
    where $k_n = \pi n / L_{\mathrm{p}}$ is the longitudinal wave number, $\phi_{\mathrm{rw}}(t)$ the amplitude, and $I_{m_{\mathrm{rw}}} $ is the modified Bessel function of the first kind and order $m_{\mathrm{rw}}$.

\section{Summary of linearized theory \label{sec_lin_theory}}
    Theoretical work on the compression by rotating electrostatic fields has been done in the linear theory framework \cite{Eggleston1999,Kiwamoto2005, Lazarow_2023}, which studies the response of a system to small perturbations to the equilibrium state. The equilibrium distribution is given by $f_0 = n_0(r) f_{\mathrm{v},0}(v)$ and the small perturbations of the distribution function and the electric potential $\phi$ are longitudinal and azimuthal Fourier decomposed into: 
    \begin{equation}
        \begin{split}
        \delta f= \sum_{nm} \delta f_{nm}(r,v) \mathrm{e}^{-i(k_n z + m\theta)}e^{p_{nm}t} \\
        \delta \phi=\sum_{nm} \delta \phi_{nm}(r) \mathrm{e}^{-i(k_n z + m\theta)}e^{p_{nm}t} 
        \end{split}
        \label{equ: decomposition}
    \end{equation}
    where $p_{nm} = i \omega_{nm} + \gamma_{nm}$ is the corresponding complex frequency with the damping factor $\gamma_{nm}$. Plugging these relations into the Vlasov equation gives the relation between the coefficients:
    \begin{equation}
        \begin{split}
            \delta f_{nm}(r,v) = \\\frac{ -\frac{q}{m_{\mathrm{q}}}ik_n n_0(r) \partial_v f_{\mathrm{v},0} - \frac{im}{B_0r}\partial_r n_0(r)f_{\mathrm{v},0}}{p- im\omega_{\mathrm{D}}-ivk} \delta \phi_{nm}(r)
        \end{split}
        \label{equ: f_nm - phi_nm}
    \end{equation}
    The second term on the right-hand side of Eq. \ref{equ: f_nm - phi_nm} is called diamagnetic drift. For a flat-top profile, $\partial_r n_0 =-n_0 \delta(r -r_{\mathrm{p}})$, so that the diamagnetic term vanishes inside the plasma column \cite{Driscoll1988} but contributes a surface term at the plasma edge. Since this term is proportional to m, it is absent for azimuthally symmetric modes. It is neglected in the following derivation of the dispersion relation and of the response to the rotating wall drive. For the full dispersion relation of a cold plasma, see \cite{Davidson}. 
    
    To get the dispersion relation, Poisson's equation in first order has to be solved together with Eq. \ref{equ: f_nm - phi_nm}. For every $(n,m)$ there is an infinite number of solutions, described by the radial mode number $l$. The radial dependency $g_{nml}(r)$ of eigenmode $(n,m,l)$ inside and outside the plasma column is then given by \cite{Davidson}:
    
    \begin{subequations}
        \begin{equation}
            \begin{split}
                g_{nml}(r) =     J_m(a_{nml} r) \\
                \text{(inside)}
            \end{split} 
            \label{equ: g(r)_inside}
        \end{equation}
        \begin{equation}
            \begin{split}
                 g_{nml}(r) = J_m(a_{nml} r_{\mathrm{p}}) \\
                    \times \frac{I_m(k_n r) K_m(k_n r_{\mathrm{w}}) - K_m(k_n r) I_m(k_n r_{\mathrm{w}})}{I_m(k_n r_{\mathrm{p}}) K_m(k_n r_{\mathrm{w}}) - K_m(k_n r_{\mathrm{p}}) I_m(k_n r_{\mathrm{w}})}   \\
                        \text{(outside)}
            \end{split}
            \label{equ: g(r)_outside}
        \end{equation}
        \label{equ: radial_dependency}
    \end{subequations}
    
    This defines the transverse wave number $a_{nml}$, which is calculated via\cite{Davidson, wave_launching}:
    \begin{equation}
        \begin{split}
            \frac{a_{nml} J_{m+1}(a_{nml} r_{\mathrm{p}}) }{J_m(a_{nml} r_{\mathrm{p}})} \\= k_n
            \frac{I_{m+1}(k_n r_{\mathrm{p}})K_m(k_n r_{\mathrm{w}}) - K_{m+1}(k_n r_{\mathrm{p}})I_m(k_n r_{\mathrm{w}})}{I_m(k_n r_{\mathrm{p}})K_m(k_n r_{\mathrm{w}}) - K_m(k_nr_{\mathrm{p}})I_m(k_n r_{\mathrm{w}})}
        \end{split}
    \end{equation}
    
    For a low damping factor and a Maxwellian velocity distribution, the corresponding eigenmode frequencies $\omega_{nml}$ are given by: 
    \begin{equation}
        \Omega^2 
        \simeq \omega_{\mathrm{p},nml}^2  \left( 1 +  \frac{3 v_{\mathrm{th}}^2 k_n^2}{\omega_{\mathrm{p},nml}^2}\right)
        \label{equ: 3D_dispersion_relation}
    \end{equation}
    with $\Omega = \omega_{nml} - m\omega_{\mathrm{D}}$, the longitudinal wave number $k_n = 2 \pi n/\mathcal{L}$, the thermal velocity $v_{\mathrm{th}}$, and $\omega_{\mathrm{p},nml}^2 = \omega_{\mathrm{p}}^2 k_n^2/(a_{nml}^2 + k_n^2)$.
     Its corresponding linear damping factor $\gamma_{nml}$ is given by \cite{Hollmann2000}:
    \begin{equation}
        \gamma_{nml} \simeq - \sqrt{\frac{\pi}{8}} (\omega_{nml}- m\omega_{\mathrm{D}}) \frac{v_{\mathrm{ph}}^3}{v_{\mathrm{th}}^3}e^{-v_{\mathrm{ph}}^2 / 2v^2_{\mathrm{th}}} 
    \end{equation}
    with the phase velocity $v_{\mathrm{ph}}=(\omega_{nml}- m\omega_{\mathrm{D}})/k_n$.
    
     For pure diocotron modes with longitudinal wave number $k_n=0$, all eigenfrequencies $\Omega_{nml}$ would be zero and $\omega_{nml} = m \omega_{\mathrm{D}}$. But for $k_n=0$, the diamagnetic term in Eq. \ref{equ: f_nm - phi_nm} cannot be neglected anymore, and the dispersion relation for these so called flute modes becomes \cite{Davidson}:
     \begin{equation}
        \omega_{m}=\omega_{\mathrm{D}}\left(m+\left( \frac{r_{\mathrm{p}}}{r_{\mathrm{w}}}\right)^{2m} -1 \right)
        \label{equ: 2D_dispersion_relation}
     \end{equation}

\paragraph{Response to rotating wall drive} 
    In order to use the same decomposition as before, the rotating wall field is projected onto the eigenmodes:
    \begin{equation}
       V_{nml} = \frac{\int_0^{2L}\mathrm{d}z \int_0^{2\pi}\mathrm{d}\theta\int_0^{r_{\mathrm{w}}}\mathrm{d}r ~r ~\Psi_{nml} \Phi_{\mathrm{RW}}^{\ast}}{\int_0^{2L}\mathrm{d}z \int_0^{2\pi}\mathrm{d}\theta\int_0^{r_{\mathrm{w}}}\mathrm{d}r ~r ~\Psi_{nml} \Psi_{nml}^{\ast}}
    \end{equation}
    where $\Psi_{nml}=g_{nml}(r) \cos{(k_n z)} e^{-im\theta}$ represents the spatial decomposition of the eigenmodes of the finite plasma column.
    Using the orthogonality of Fourier series, the projection  can be written as
    \begin{equation}
        V_{nml} = V_m(t)C_{nml}
    \end{equation}
    where
    \begin{equation}
         V_m(t)= \phi_{\mathrm{rw}}(t)e^{-i\frac{m}{m_{\mathrm{rw}}}\omega_{\mathrm{rw}} t}
    \end{equation}
        \begin{equation}
        \begin{split}
          C_{nml} = \delta_{|m|,|m_{\mathrm{rw}}|} \frac{1}{\pi n}(-1)^{n} \sin{(k_n \Delta z)}   \\ \times  \frac{
           \int_0^{r_{\mathrm{p}}}~\mathrm{d}r ~r ~g_{nm_{\mathrm{rw}}l}(r)
           \frac{I_{m_{\mathrm{rw}}}(k_n r)}{I_{m_{\mathrm{rw}}} (k_n r_{\mathrm{w}})} }{\int_0^{r_{\mathrm{p}}}~\mathrm{d}r  ~r ~|g_{nm_{\mathrm{rw}}l}(r)|^2 } ~~.
        \end{split}
    \end{equation}
    For a plasma, which is initially in equilibrium and an external rotating wall drive as described above, the analysis in \cite{wave_launching} can be expanded to the azimuthal non-symmetric case. The amplitude evolution of the driven eigenmodes is given through the convolution:
    \begin{equation}
        \begin{split}
            \delta \phi_{nml}(t) = -(\omega_{nml} - m\omega_{\mathrm{D}})  C_{nml} \\ \int_0^t \mathrm{d}\tau\left[  \phi_{\mathrm{rw}}(\tau) e^{-im(\frac{\omega_{\mathrm{rw}}}{m_{\mathrm{rw}}} - \omega_{\mathrm{D}} )\tau} \right.   \\ \left.
            \times \sin{((\omega_{nml} - m\omega_{\mathrm{D}})(t-\tau))} e^{\gamma_{nml} (t-\tau)}\right] ~~.
        \end{split}
        \label{equ: mode_evolution}
    \end{equation}

    From Eq. \ref{equ: mode_evolution} a transient oscillation of the mode amplitude with the beating frequency $\omega_{\mathrm{rw}} - \omega_{nml}$ is expected, which will be damped to a constant value of the mode amplitude. The final amplitude after the transient oscillation depends on the damping factor $\gamma$ and the frequency slip between $\omega_{\mathrm{rw}}$ and $\omega_{nml}$. The smaller these quantities are, the higher will be the final eigenmode amplitude. For more details on the derivation of Eq. \ref{equ: mode_evolution}, see appendix \ref{sec_derivation_amp_excitation}.

\section{Modes from simulation noise spectra \label{sec_noise_spectra}}
\begin{figure}
        \centering
        \includegraphics[]{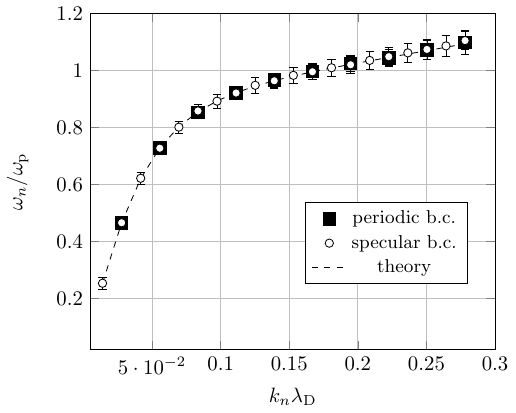}
        \caption{Dispersion relation longitudinal modes from noise (1D Simulation). Finite plasma has additional modes at half wavelengths of the infinite plasma column.}
        \label{fig: 1D-dispersion_relation}
    \end{figure}

\begin{figure}
        \centering
        \includegraphics[]{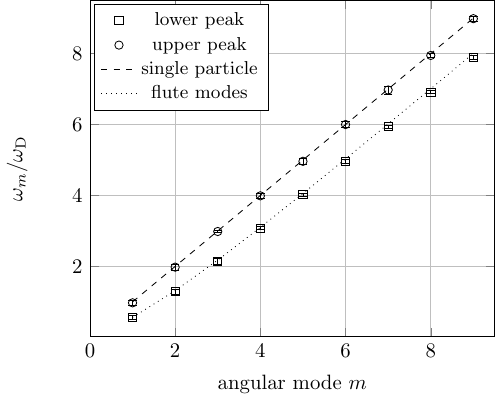}
        \caption{Dispersion relation transversal modes from noise (2D Simulation). The lower branch corresponds to the flute perturbations, while the upper branch corresponds to the single particle rotation frequency.
        }
        \label{fig: 2D-dispersion_relation}
    \end{figure}

    The macro-particle positions are initialized by sampling the radial coordinate from a probability distribution proportional to $r$, ensuring a uniform, area-weighted density within the plasma column. The azimuthal and axial coordinates are drawn uniformly from $[0,2\pi]$ and $[0,L_{\mathrm{p}}]$, respectively, so that the resulting ensemble corresponds to a homogeneous cylindrical density distribution in the domain $[0,r_{\mathrm{p}}]\times[0,2\pi] \times [0,L_{\mathrm{p}}]$. Velocities are drawn from a Maxwellian distribution, centered around zero and with thermal velocity $v_{\mathrm{th}}$. The finite number of randomly initialized macro-particles introduces Schottky (shot) noise in the space charge field, i.e. statistical fluctuations of the mode amplitudes. Throughout this paper, “noise” refers to these fluctuations. The frequencies of the eigenmodes are obtained from the spectrum of these fluctuations by an FFT of the complex mode amplitudes. The uncertainty bars in Figs. \ref{fig: 1D-dispersion_relation}, \ref{fig: 2D-dispersion_relation} and \ref{fig: 3D-dispersion_relation} denote the full width at half maximum of the peaks, which includes the damping of the modes and the finite observation time and is an upper bound for the uncertainty of the peak position. In five runs with different random initial distributions, the peak positions agreed within the spectral resolution $\Delta \omega = 2\pi/T$.

    First, the spectral macro-particle scheme is applied to one-dimensional longitudinal dynamics. The reduction is done by setting the angular mode to zero $m=0$ and approximating the radial profile by $J_0(a_{01} r)$, the lowest radial mode $l=1$. The parameters for all simulations presented in this paper are summarized in table \ref{tab: simu_parameter}. In Fig. \ref{fig: 1D-dispersion_relation} the results for the mode frequencies are presented, for the case of periodic and specular boundary conditions, where the finite plasma has additional modes at half wavelengths of the infinite plasma. In both cases, the results are in agreement with Eq. \ref{equ: 3D_dispersion_relation}.

\begin{figure}
        \centering
        \includegraphics[]{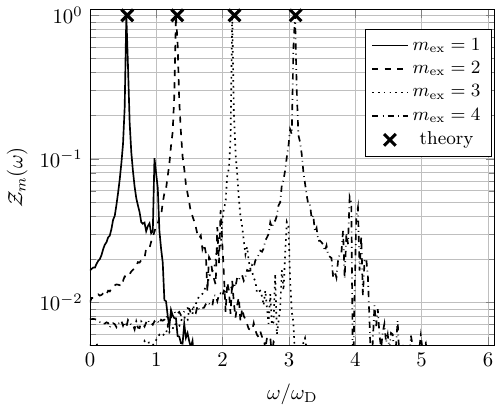}
        \caption{Response functions (normalized) to a broadband excitation with multipole order $m_{\mathrm{ex}}$. Only the corresponding flute modes are excited, which are the eigen modes of the system.}
        \label{fig: 2D_response_fct}
    \end{figure}

\begin{figure}
        \centering
        \includegraphics[]{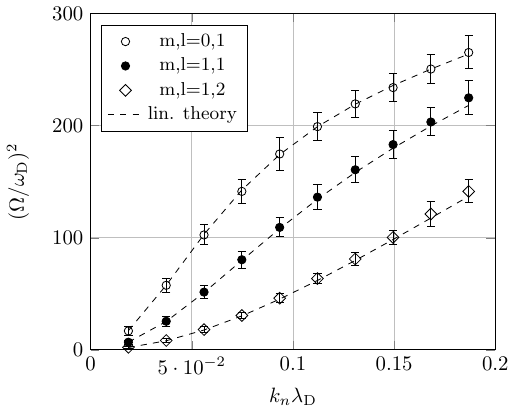}
        \caption{Dispersion relation of finite plasma from noise (3D Simulation). Results for angular and radial modes m,l=0,1 ($\circ$), m,l=1,1 ($\bullet$) and m,l=1,2 ($ \diamond $). The corresponding results from Eq. \ref{equ: 3D_dispersion_relation} are indicated by dashed lines. }%
        \label{fig: 3D-dispersion_relation}
    \end{figure}

    Another test case are the pure transversal modes, where $k_n=0$ and flute modes according to Eq. \ref{equ: 2D_dispersion_relation} are expected. In Fig. \ref{fig: 2D-dispersion_relation} the results from the 2D noise spectra for $r_{\mathrm{p}}/r_{\mathrm{w}}=0.75$ are shown. For every angular mode, there are two distinct peaks. The lower peaks correspond to the flute modes of Eq. \ref{equ: 2D_dispersion_relation}. 
    To test whether both peaks correspond to eigenmodes, an external field $\phi_{\mathrm{ext}} = \phi_0 (r/r_{\mathrm{w}})^{m_{\mathrm{ext}}} \cos{(m_{\mathrm{ext}} \theta - \omega t)}$ with multipole order $m_{\mathrm{ext}}$ and frequencies uniformly spread over $\omega \in [0,10\,\omega_{\mathrm{D}}]$ is applied. The linear response function $\mathcal{Z}_m(\omega) = |\phi_m(\omega)|/| \phi_{\mathrm{ext}}(\omega)|$, see Fig. \ref{fig: 2D_response_fct}, has maxima at the eigenfrequencies of the system. Only the lower peak of the angular mode  $m= m_{\mathrm{ext}}$ responds, whereas the height of the upper peak remains unchanged. The lower peaks are, therefore, the diocotron (flute) eigenmodes. The upper  peaks at $\omega = m \omega_{\mathrm{D}}$ are not collective modes. In a flat-top column all guiding centers rotate rigidly with $\omega_{\mathrm{D}}$, so that the incoherent contribution of each macro-particle to the $m$-th azimuthal coefficient oscillates at $m\omega_{\mathrm{D}}$. This discreteness noise is advected by the $E \times B$ flow and, lacking a restoring force, shows no resonant response.

    For the full three-dimensional dispersion relation, the simulation was performed with specular boundary conditions. The results for the angular and radial mode numbers (0,1), (1,1) and (1,2) are shown in Fig. \ref{fig: 3D-dispersion_relation}. The expected frequencies from linear theory are close to the eigen frequencies obtained from the simulation, the expected frequencies deviate less than 1.5~\% for $m=0$ and less than 6.5~\% for m=1 in $(\Omega/\omega_{\mathrm{D}})^2$. This demonstrates, that also higher angular and radial modes are resolved correctly by the simulation.

    In longitudinal, transversal and full 3D space, the spectral macroparticle simulation scheme is capable of retrieving the correct eigen frequencies from noise. Moreover, the spectral decomposition of the space charge solver allows direct tracking of the eigenmodes. This enables us to study the response of individual eigenmodes to external perturbations, which will be discussed in the next section.

    \begin{figure}
         \centering
        \includegraphics[]{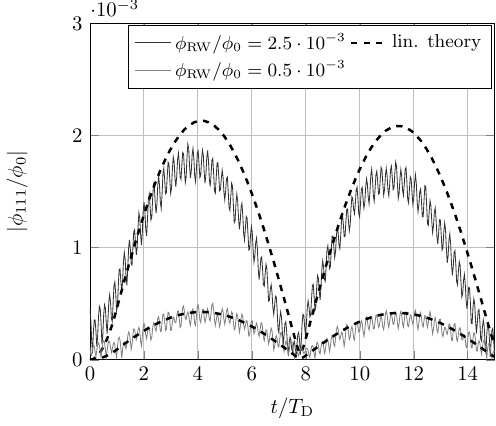}
        \caption{ \textit{Response to rotating wall drive -} temporal evolution of mode amplitude (1,1,1) for a drive with $\omega_{\mathrm{rw}} \approx 1.04 \omega_{1,1,1}$, $m_{\mathrm{rw}} =1$ and two different amplitudes. There is a beating with frequency $(\omega_{\mathrm{rw}}-\omega_{nml})$ and a small additional jitter with frequency $(\omega_{\mathrm{rw}} + \omega_{nml}-2m\omega_{D})$ in simulation data. The deviations can be associated with the diamagnetic term.}
        \label{fig: mode_amp_vs_time}
    \end{figure}

\section{Plasma Response to Rotating Wall Drive \label{sec_RW_response}}
    For simplicity, a perfect rotating wall drive is assumed, where the segmented nature of a real rotating wall electrode is neglected. The angular mode number of the rotating wall drive is set to $m_{\mathrm{D}} =1$ and its amplitude increases from zero to its final value during one plasma rotation $T_{\mathrm{D}}= 2 \pi / \omega_{\mathrm{D}}$ at a constant rate. 

    The response of the eigenmode amplitudes to the rotating wall drive is studied in Fig. \ref{fig: mode_amp_vs_time}, where the drive frequency was set to $\omega_{\mathrm{rw}} \approx 1.04 \omega_{111}$. The temporal evolution of eigenmode $\phi_{111}$ is shown for two different drive amplitudes. For comparison with linear theory, Eq. \ref{equ: mode_evolution} was calculated with damping factor $\gamma_{nml}= 0.001\omega_{D}$. Note that in the derivation of Eq. \ref{equ: mode_evolution} zero initial perturbation was assumed, which is not the case in the simulation, where an initial noise is introduced by the macro-particle initialization. However, Eq. \ref{equ: mode_evolution} is capable of describing the features of the eigenmode evolution in the linear region.

    From Eq. \ref{equ: mode_evolution} a beating with frequency $\omega_{\mathrm{rw}}-\omega_{nml}$ is expected and is also present in the simulation. Besides, there is also a small additional jitter with frequency $(\omega_{\mathrm{rw}} + \omega_{nml}-2m\omega_{D})$ on the simulation signal. This jitter originates from the non-zero initial perturbations. In simulation and from Eq. \ref{equ: mode_evolution} there is a damping of the initial transient oscillation of the amplitude around its mean value. With time, the kinetic energy increases, which in turn changes the equilibrium plasma parameter. The eigenfrequencies are shifted to higher values and the damping factor changes too. Therefore, over time the beating frequency and the mean value of the eigenmode amplitude shift.

    \begin{figure}
        \centering
        \includegraphics[]{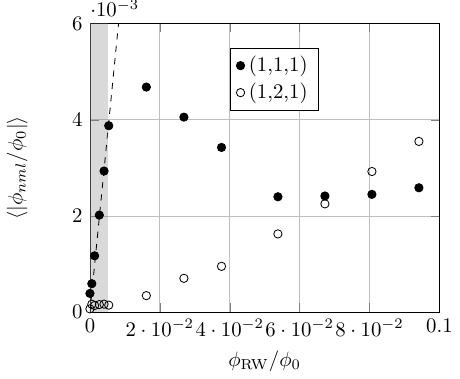}
        \caption{ \textit{Response to rotating wall drive-} Mean amplitude of excited modes for drive with $m_{\mathrm{rw}} =1$ and tuned close to eigen frequency of (1,1,1). ($\bullet$) mode (1,1,1). As given by Eq.\ref{equ: mode_evolution} and indicated by the dashed line the resulting amplitude should be proportional to the RW amplitude. This is the case until $\phi_{\mathrm{RW}}/\phi_0 = 0.005$. ($\circ$) mode (1,2,1) is initially close to zero but starts to rise above $\phi_{\mathrm{RW}}/\phi_0 = 0.005$. Above this value the response is not linear anymore and other modes are getting excited.}
        \label{fig: mode_amp_vs_RW_amplitudes}
    \end{figure}

    In Fig. \ref{fig: mode_amp_vs_RW_amplitudes} the mean values of the eigenmodes (1,1,1) and (1,2,1) are plotted against the drive amplitude. The drive frequency is tuned close to the eigen frequency of (1,1,1) and the mean value is taken for a simulation time of 8 $T_{\mathrm{D}}$. For small amplitudes, up to $\phi_{\mathrm{RW}}/\phi_0 = 0.005$, $\langle |\phi_{111}|\rangle$ is proportional to the drive amplitude, and $\langle |\phi_{121}|\rangle$ does not change. However, above this drive amplitude, $\langle |\phi_{111}|\rangle$ is no longer proportional to the drive amplitude, and $\langle |\phi_{121}|\rangle$ starts to increase. 
    The temporal spectrum of the $(1,2,1)$ coefficient shows a peak near its natural eigenfrequency $\omega_{121}$ and no component near $2\omega_{111}$. The same is found for a drive restricted to the longitudinal mode number $n_{\mathrm{rw}}=\,1$. The heating shifts $\omega_{121}$, which lies above $\omega_{\mathrm{rw}}$, further away from $\omega_{\mathrm{rw}}$.
    
    \begin{figure}
        \centering
        \includegraphics[]{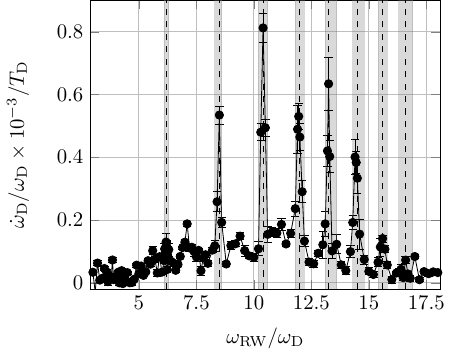}
        \caption{ \textit{Response to rotating wall drive-} Change of the mean rotation frequency depending on the drive frequency for $\phi_{\mathrm{RW}}/\phi_0 = 3 \cdot 10^{-4}$ ($\bullet$). Dashed lines show the eigen frequencies $\omega_{n11}$, obtained from the simulation. When an eigenmode is excited, there is an enhanced compression rate. The height of the peaks is connected to the mode spectrum of RW drive, which depend on the geometry. The small peak around $7\omega_{\mathrm{D}}$ is connected to the second radial eigenmode n=4. The contribution of the higher radial modes become apparent in the background compression rate.}
        \label{fig: compression_rate_low_phi_RW}
    \end{figure} 

    Without external drive, the kinetic energy of the macro-particles first increases by about $0.8\,\%$ during an initial phase of about $0.5\,T_\mathrm{D}$, caused by the random initialization, and changes slowly afterwards by about 0.01 \% per $T_\mathrm{D}$ for $N_\mathrm{p} =\, 2\cdot10^5$ (Fig. \ref{fig: E_kin_vs_time}, Appendix A). The heating observed with drive is larger by orders of magnitude and occurs only close to the eigenfrequencies. It can be thereby attributed to Landau damping of the excited modes. For the parameters of Fig. 7 the response is linear up to $\phi_{\mathrm{RW}}/\phi_0 = 0.005$ within $8\, T_\mathrm{D}$ and system can be considered to be in the weak drive regime.

    To investigate the compression behavior, the change of the mean rotation frequency was tracked in a frequency range around the eigenmodes. For radial profiles close to a flat top profile, the mean rotation frequency is proportional to the density, $\omega_{\mathrm{D}}=qn_0/ 2\epsilon_0B_0$. As the compression rate depends on $v_{\mathrm{ph}}/v_{\mathrm{th}}$ \cite{Eggleston1999,Kiwamoto2005,Lazarow_2023}, the parameters were chosen as a trade off between high compression rates and low damping of the eigenmodes.

    \begin{figure}
        \includegraphics[]{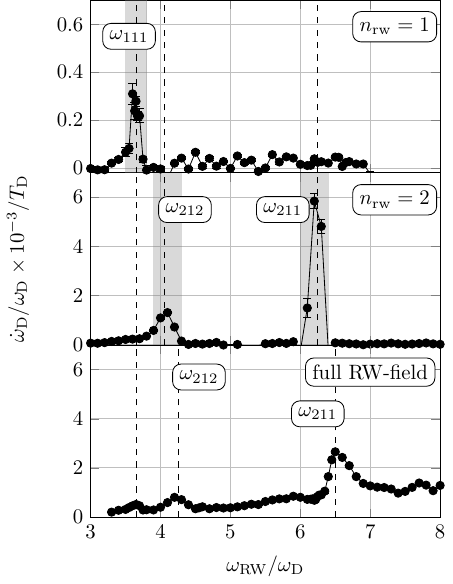}
        \caption{ \textit{Response to rotating wall drive-} Change of the mean rotation frequency depending on the drive frequency for $\phi_{\mathrm{RW}}/\phi_0 = 3 \cdot 10^{-3}$ ($\bullet$). If only one specific longitudinal mode is allowed for the drive, only modes corresponding to this longitudinal mode are excited. In case of multiple longitudinal modes from the drive, there is a background compression due to the higher radial modes. Additionally, the system heats up during the simulation run. This shifts the eigenfrequencies and decreases the eigenmode response to the external drive at a fixed frequency.}
        \label{fig: compression_rate}
    \end{figure}

    In Fig. \ref{fig: compression_rate_low_phi_RW}, the drive amplitude was set to $\phi_{\mathrm{RW}}/\phi_0 = 3 \cdot 10^{-4}$, which is in the linear regime as identified in Fig. \ref{fig: mode_amp_vs_RW_amplitudes}. In addition, rotating wall drive modes with higher longitudinal mode numbers were enhanced to enable investigation of their influence. The change in the mean parallel kinetic energy,$E_{\parallel} = \sum_j m v_{z,j}^2 /2$, was up to \SI{80}{\percent} for frequencies close to an eigen-frequency and considerably less otherwise. The peaks in the simulation data can be associated with specific eigenmodes by comparing the eigenmode amplitudes. At each peak, there is only one eigenmode that is notably excited. This is expected for the weak drive regime. The relative heights of the peaks are connected to the mode spectrum of the rotating wall drive, which in turn depends on the geometry of the rotating wall electrode, the trap and the trapped plasma. This geometry determines how strongly specific eigenmodes are excited by the drive. The marked peaks belong to the longitudinal mode numbers from $n=2$ to $n=9$, angular mode $m=1$ and radial mode $l=1$. In addition to the first radial mode, there is also a peak at $7.1\omega_{\mathrm{D}}$, which can be associated with the eigenmode (4,1,2). The other higher radial modes contribute to a small background compression rate. If only a specific longitudinal mode of the rotating wall drive is turned on, only modes according to this mode number are excited. In this case, compression is observable only around the corresponding frequencies, as shown in Fig. \ref{fig: compression_rate} (top and middle). For these simulation runs, the drive amplitude was set to a higher amplitude of $\phi_{\mathrm{RW}}/\phi_0 = 3 \cdot 10^{-3}$.

    During the simulation time, the mean square radius decreased by $0.1\,\%$ ($\varphi_{\mathrm{RW}}/\varphi_0 = 3\cdot10^{-4}$) and $1\,\%$ ($\varphi_{\mathrm{RW}}/\varphi_0 = 3\cdot10^{-3}$) on resonance, consistent within $3\,\%$ with the relative change of $\omega_\mathrm{D} \propto 1/\langle r^2\rangle$. The plasma edge moved by less than $1\,\%$ of the wall radius. For $\varphi_{\mathrm{RW}}/\varphi_0 = 3\cdot10^{-3}$, the density close to the axis ($r < 0.05\,r_{\mathrm{p}}$) increased at the peaks of the compression rate at $\omega_{211}$ by up to a factor of $2$ for $n_{rw} = 2$ only and $1.4$ for the full drive spectrum; elsewhere the profile stayed close to the initial flat-top profile.

    However, as shown in Fig. \ref{fig: compression_rate} (bottom), if multiple longitudinal modes of the rotating wall drive are turned on, there can be considerable heating during the simulation time.  Here, heating refers to an increase of the parallel temperature; in the drift-kinetic model, the perpendicular energy is not a dynamical variable. The heating shifts the eigenfrequencies upward, Eq. \eqref{equ: 3D_dispersion_relation}, so that a drive at fixed frequency becomes detuned and the excited amplitudes and the compression rate decrease. This points to the limits of the model at hand. There are no cooling mechanisms included and external excitation will heat up the plasma, while in reality cooling mechanisms, like cyclotron cooling, will limit or even counterbalance the heating introduced by the external wall drive. This sets the limits of applicability for simulations of external drives to short term responses and low drive amplitudes. Thus, with the current simple model of trapped plasma, it is not possible to simulate long term compression behavior and to investigate the strong drive regime.

\section{Conclusion}
    Our study demonstrates the capability of the spectral macroparticle simulation scheme to accurately retrieve the dispersion relation of a trapped plasma from simulation noise and to track the evolution of eigenmodes and their response to external excitations. We have adapted an existing expression for the excitation of eigenmodes due to an external drive from the azimuthally symmetric to the azimuthally nonsymmetric case and verified its validity through direct comparison with our simulation results. We identified the limit of the linear response with increasing drive amplitude and the onset of mode mixing. Moreover, we have successfully simulated the compression of a plasma by a rotating wall drive in the weak drive regime. These achievements highlight the potential of our simulation method in providing a comprehensive understanding of plasma behavior under various settings of the plasma properties, trap geometry, and rotating wall drive parameter. 

    Our full Vlasov simulation model goes beyond the known linearized theory and allows us to predict the limits of linear, single mode response. This is also the the expected transition to the strong drive regime. However, our model does not include cooling mechanisms, resulting in heating due to the external drive. This limits the simulation time and amplitude of the applied drive. Future work will include a diffusion/cooling term to enable studies of the transition regime.

\begin{acknowledgments}
    Part of this work was supported by the Federal Ministry of Education and Research (BMBF) 05P24RD4. The authors gratefully acknowledge the computing time provided to them on the high-performance computer Lichtenberg II at TU Darmstadt, funded by the German Federal Ministry of Research, Technology and Space (BMFTR) and the State of Hesse.
\end{acknowledgments}

\section*{Author Declarations}
    \subsection*{Conflict of Interest}
        The authors have no conflicts to disclose.
    \subsection*{Author Contributions}
        \textbf{L. Riik:} Data curation(lead); Investigation (lead), Software(lead); Visualization (lead); Writing - original draft (lead); Formal analysis (lead); Methodology (equal); Conceptualization (equal); Writing – review \& editing (equal). 
        \textbf{O. Boine-Frankenheim:} Project Administration(lead); Supervision(lead); Methodology (equal); Conceptualization (equal); Writing - original draft (supporting); Writing – review \& editing (equal)

\section*{Data Availability Statement}
    The data that support the findings of this study are available from the corresponding author upon reasonable request. The simulation code used to generate the data is openly available at https://doi.org/10.48328/tudatalib-2320.

\appendix

\section{Spectral Macro-particle Simulation Model \label{sec_simulation_model}}
    The simulation scheme solves Poisson's equation for the space charge field of the macro-particles inside a cylindrical perfect conducting tube and pushes the drift centers according to the drift kinetic Vlasov equation \ref{equ: Vlasov_equation}. The space charge field is calculated with the help of a Fourier-Bessel decomposition. The potential and charge density are assumed to have periodic boundary conditions in longitudinal direction with periodicity length $L$ and Dirichlet b.c. $\phi(r=r_{\mathrm{w}}) = 0$ in radial direction. Therefore, they are expanded in terms of Fourier-Bessel series in cylindrical coordinates:
    \begin{align}
        \rho = \sum_{nml} \rho^{nml} J_m(\chi_{ml}r) \exp{\{-i(m\theta + k_nz)}\} \label{equ:Fourier_Bessel_series_rho}\\
        \phi = \sum_{nml} \phi^{nml} J_m(\chi_{ml}r) \exp{\{-i(m\theta + k_nz)\}} 
    \label{equ:Fourier_Bessel_series}
    \end{align}
    with $k_n=2\pi n /L$ and $\chi_{ml}$ given by the l'th zero of $J_m(\chi_{ml} r_{\mathrm{w}})$. Inserting these equations into the cylindrical Poisson's equation gives the relation between the coefficients of $\rho$ and $\phi$:
    \begin{equation}
        \left(\chi_{ml}^2 + k_n^2 \right) \phi^{nml} = \frac{\rho^{nml}}{\epsilon_0}
    \end{equation}

    For a delta distribution of macro-particles with charge $q_j$, the coefficients of the space charge potential are then given by the sum over all $N_{\mathrm{p}}$ macroparticles:
    \begin{equation}
        \begin{split}
          \phi^{nml} = \frac{1}{\epsilon_0 \left(\chi_{ml}^2 + k_n^2 \right)} \frac{1}{\pi a^2 (J_{m+1}(\chi_{ml}))^2 L} \\ \times \sum_j^{N_{\mathrm{p}}} q_j  e^{im\theta_j} e^{ik_nz_j}J_m(\chi_{ml} r_j)
        \end{split}
    \end{equation}

    Finally, the electric field is calculated from the potential $\nabla \mathbf{E}=-\phi$.
    
    \begin{figure}[h]
        \centering
        \subfloat[$E_r (r)$ for infinite cylinder with $r_{\mathrm{p}}$=0.5 $r_{\mathrm{w}}$ \label{fig: E_r_field_calc}]{\includegraphics[]{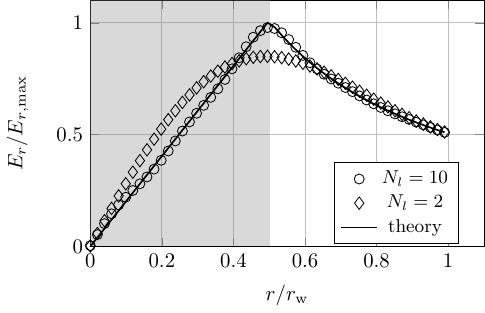}} \hfill
        \subfloat[Convergence study for fixed $N_n=10$, $N_m=10$ \label{fig: E_r_conv}]{\includegraphics[]{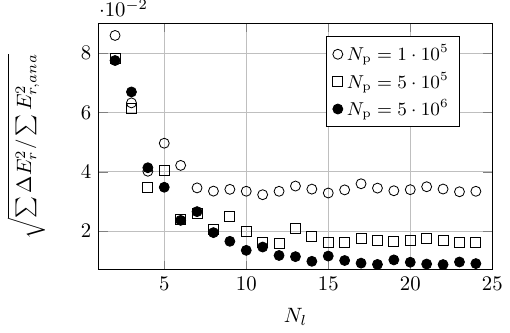}}
        \caption{(a) Radial electric field $E_r (r)$ for an infinite plasma column with $r_{\mathrm{p}}$=0.5 $r_{\mathrm{w}}$ for $N_{\mathrm{p}}=5\cdot10^5$ macro-particles. The field was calculated by the spectral space charge solver in case of ($\circ$) $N_n,N_m,N_l = 10,10,10$  and ($\diamond$) $N_n,N_m,N_l = 10,10,2 $. The gray area indicates the radial extend of the plasma. (b) Convergence study for fixed $N_n=10$, $N_m=10$. The root mean square error decreases with $N_l$ to a value limited by the number of macro-particles. In order to resolve small structures, the truncation order has to be increased together with the macro-particle number.} 
        \label{fig: field_calc}
    \end{figure}
    In Fig. \eqref{fig: E_r_field_calc}  the radial electric field is shown for the case of an infinitely long plasma column with a flat top profile. The radial extent of the column is $r_{\mathrm{p}}$ =~0.5~$r_{\mathrm{w}}$, as indicated by the gray area. In case of a truncation order of $(N_n,N_m,N_l) = (10,10,2) $, the sharp edge of the potential at the plasma radius is not properly resolved. Increasing $N_l$ increases the resolution in the radial direction. In Fig. \eqref{fig: E_r_conv} a convergence study for fixed $N_n=10$, $N_m=10$ is shown. The root mean square error decreases with $N_l$ to a value conditioned by the number of macro-particles. To increase the resolution of the space charge field calculation, the number of macro-particles has to be increased as well. So, Spectral Methods are a good choice for smooth solutions, but are limited in the case of discontinuities \cite{Numerical_Recipes}, as a high truncation order and higher particle numbers would be necessary.

    \begin{figure}[h]
        \centering
    \includegraphics[]{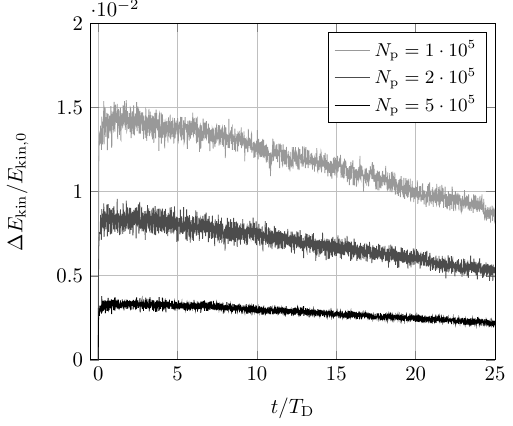}
    \caption{Evolution of the kinetic energy with time. Due to the finite macro-particle number, initial perturbations are introduced. Consequently, the kinetic energy of the macro-particle cloud increases during an initial phase.} 
     \label{fig: E_kin_vs_time}
    \end{figure}

    The macro-particles are pushed according to Eq. \eqref{equ: Vlasov_equation} with a Strang time splitting scheme. This splitting scheme has second order accuracy, as was shown in \cite{Grandgirard2006} and includes the calculation of the self field two times per time step and is momentum conserving. The simulation code is publicly accessible \cite{simulation_code}. 

    The initial phase space distribution is generated from an area weighted uniform random distribution in space and a random Maxwellian distribution in velocity, introducing initial perturbations. Consequently, the kinetic energy of the macro-particle cloud increases depending on the number of macro-particles, as shown in Fig. \ref{fig: E_kin_vs_time}, e.g. by $0.8\, \%$ for $N_\mathrm{p}=2\cdot10^5$, during a initial phase of about $0.5\, T_\mathrm{D}$. After the initial increase, the change in the parallel kinetic energy is about $0.01\,\%$ per $T_\mathrm{D}$ for $N_\mathrm{p}=2\cdot10^5$.
    
    The finite number of macro-particles and the artificial collisions cause noise and diffusion. By truncating the Fourier-Bessel series, the spectral space charge solver acts as a low-pass filter for the spatial frequencies. The cost of the field calculation scales as $\mathcal{O}(N_n \times N_m \times N_l \times N_{\mathrm{p}})$, compared with $\mathcal{O}(N_{\mathrm{p}})$ plus, e.g., $\mathcal{O}(N_{\mathrm{g}} \log N_{\mathrm{g}})$ for an FFT-based field solver in a grid-based particle-in-cell (PIC) scheme. The spectral scheme is therefore more expensive per macro-particle, but it represents the cylindrical conducting boundary exactly, is free of grid heating, filters unphysical small-scale noise and provides direct access to the eigenmode amplitudes. 
    
    The longitudinal and azimuthal eigenmode structure of the plasma column can be read out directly from the space-charge solver, while the radial eigenmode amplitudes follow from a projection onto the radial eigenfunctions 

    \begin{equation}
        c_{nm,ll'} = \frac{\int_0^{r_w} r\, g_{nml}(r)\, J_m(\chi_{ml'} r)\, \mathrm{d}r}{\tfrac{1}{2}\, r_w^2\, J_{m+1}^2(\chi_{ml'} r_w)} ~~.
        \label{equ: radial_coeff_matrix}
    \end{equation}
    The eigenfrequencies are not affected by the projection, since all coefficients of a given $(n,m)$ contain the same spectral lines with different weights. For the geometry used in the present work the coefficient matrix is given in table \ref{tab: coefficient matrix}. In case of $l~=~1$ the TG mode is dominated by the first radial mode of the spectral solver.
    
    \begin{table}[]
        \centering
        \begin{tabular}{|c|c|c|c|c|}
        \hline
        $l$      &     $l'=1$     &      $l'=2$      &     $l'=3$     &      $l'=4$   \\ \hline
     1 & 1.027 & 0.201 &-0.128 & 0.070 \\
     2 &-0.073 & 0.949 & 0.511 &-0.184  \\
     3 & 0.037 &-0.180 & 0.738 & 0.801 \\
     \hline
        \end{tabular}
        \caption{Radial coefficient matrix for $r_{\mathrm{p}}/r_{\mathrm{w}}=0.75$ and $n=1$, $m=1$. For $l=1$ the TG mode is dominated by the first radial mode of the spectral solver.}
        \label{tab: coefficient matrix}
    \end{table}

    The simulation parameters for all simulation runs presented are summarized in table \ref{tab: simu_parameter}. The dimensionless representation of the longitudinal mode is $k_n \lambda_\mathrm{De}$, where $k_n=2n\pi/\mathcal{L}$ is the fundamental longitudinal wave number and $\lambda_{\mathrm{De}} = v_{\mathrm{th}}/\omega_{\mathrm{p}}$ follows from the initial thermal velocity. For $k_1 \lambda_{\mathrm{De}}=0.018$ this corresponds to $\lambda_{\mathrm{De}} \approx 0.006\, L_{\mathrm{p}} \approx 0.03\, r_{\mathrm{w}} \approx 0.04\, r_{\mathrm{p}}$ and for $k_1 \lambda_{\mathrm{De}}=0.028$ this corresponds to $\lambda_{\mathrm{De}} \approx 0.045\, r_{\mathrm{w}} \approx 0.06 \,r_{\mathrm{p}}$.

    \newcommand{\tbd}{\textcolor{red}{[\,\ldots]}}   

    \begin{table*}
        \caption{Simulation parameters. The parameters in the
        upper block are identical for all simulation runs; the parameters in the
        lower block are listed for each figure. $T_D = 2\pi/\omega_{\mathrm{D}}$ is the
        rotation period, $N_{\mathrm{p}}$ the number of macro-particles and
        $(N_n, N_m, N_l)$ the truncation orders of the longitudinal, azimuthal
        and radial expansion. }
        \begin{ruledtabular}
        \begin{tabular}{lccccc}
        \multicolumn{6}{l}{\textit{Common to all simulation runs}} \\
        $r_{\mathrm{p}}/r_{\mathrm{w}}$               & \multicolumn{5}{c}{0.75} \\
        $r_{\mathrm{w}}/L_{\mathrm{p}}$               & \multicolumn{5}{c}{0.2}  \\
        $\omega_{\mathrm{D}}/\omega_{\mathrm{p}}$     & \multicolumn{5}{c}{0.06} \\
        $\Delta t/T_{\mathrm{D}}$ (Fig. 2~$\Delta t/(2\pi/ \omega_{\mathrm{p}})$) & \multicolumn{5}{c}{0.01} \\
        \hline
        \multicolumn{6}{l}{\textit{Run-specific}} \\
        Parameter & Fig.~2 & Figs.~3--4 & Fig.~5 & Figs.~6--7 & Figs.~8--9 \\
        \hline
        Model
          & 1D, $m=0$, $J_0(a_{01}r)$ & 2D, $k_n=0$ & 3D & 3D & 3D \\
        Longitudinal b.c.
          & periodic / specular & -- & specular & specular & specular \\
        $k_1\lambda_{\mathrm{De}}$
          & 0.018 & -- & 0.018 & 0.018 & 0.028 \\
        $v_{\mathrm{th}}/v_{\mathrm{ph}}$ (resolved modes)
          & $<0.25$ & -- & $<0.4$ & $<0.25$ & 0.18--0.27 \\
        $(N_n, N_m, N_l)$
          & $(10,0,0)$/$(20,0,0)$ & $(0,10,10)$ & $(7,7,10)$ & $(5,5,5)$ & $(5,5,10)$ \\
        $N_{\mathrm{p}}$
          & $10^3$& $10^4$& $5 \cdot10^5$& $2 \cdot10^5$  & $2 \cdot10^5$ \\
        Duration$/T_{\mathrm{D}}$
          & 60 & 40 & 20 & 15 \& 8 & 6 \\
        \end{tabular}
        \end{ruledtabular}
        \label{tab: simu_parameter}
    \end{table*}

    The simulations for Fig. \ref{fig: 1D-dispersion_relation} to Fig. \ref{fig: mode_amp_vs_time}
    were performed on a standard PC. For $N_\mathrm{p}=10^5$, truncation order $(5,5,10)$, 1000 timesteps and 8 cores the computation time of the main loop was 54 minutes. The parameter studies in Fig. \ref{fig: mode_amp_vs_RW_amplitudes} to Fig. \ref{fig: compression_rate} were performed on the Lichtenberg II cluster at TU Darmstadt to parallelize the different parameter runs. A performance optimized version written in the julia programming language is now available \cite{riik-smushkin2026}. The optimized version performs the same calculation in a time of 32 seconds.

\section{Time Dependence of Excited Eigenmodes \label{sec_derivation_amp_excitation}}
    The following calculation is an adaption from \cite{wave_launching} to the azimuthal nonsymmetric case. The perturbations are described by
    \begin{equation}
    \begin{split}
        \delta f = \sum_{nm} \delta f_{nm}(r,v) \cos{(k_n z)} e^{-im\theta} e^{pt}=\delta f^+ + \delta f^- \\
        \delta \phi = \sum_{nm} \delta \phi_{nm}(r) \cos{(k_n z)} e^{-im\theta}e^{pt} =\delta \phi^+ + \delta \phi^-
    \end{split}
    \end{equation}
    where the cosine function is split into two exponential functions, in order to facilitate the calculation:
    \begin{align}
        \begin{split}
            \partial f^{\pm} = \frac{1}{2}\sum_{nm} \delta f^{\pm}_{nm}(r,v) e^{\pm ik_nz} e^{-im\theta}e^{pt} \\
            \partial \phi^{\pm} = \frac{1}{2} \sum_{nm} \delta \phi^{\pm}_{nm}(r)  e^{\pm ik_nz} e^{-im\theta} e^{pt} 
        \end{split}
    \end{align}
    
    The external potential is added to the potential perturbation:
    \begin{equation}
        \begin{split}
            \partial \phi^{\pm} = \frac{1}{2} \sum_{nm} \delta\hat{\phi}^{\pm}_{nm}(r)  e^{\pm ik_n z} e^{-im\theta} e^{pt} \\+ \frac{1}{2} \sum_{nm} V(p) C^{\pm}_{nm}(r) e^{\pm ik_n z}
        \end{split}
    \end{equation}
    
    This mode decomposition is inserted into the linearized Vlasov equation
    \begin{equation}
        \begin{split}
            \partial_t \delta f +v\partial_z \delta f - \frac{1}{Br} \left(\partial_{\theta}\delta\phi \partial_r f_0- \partial_r \phi_0 \partial_{\theta} \delta f \right) \\ - \frac{q}{m}\partial_z \delta \phi \partial_v f_0 =0
        \end{split}
    \end{equation}
    where $ f_0 = n_0(r) f_{\mathrm{v},0}(v)$, to get the coefficient relation:
    \begin{equation}
        \delta\hat{f}^{\pm}_{nm}(r,v) = \frac{\pm \frac{q}{m}ik_n \partial_v f_{\mathrm{v},0} \left(\delta \phi^{\pm}_{nm} + V(p)C^{\pm}_{nm} \right) }{p- im\omega_{\mathrm{D}}\pm ivk_n} 
        \label{equ:df_pm_launch}
    \end{equation}
    
    The linearized Poisson's equations in first order is:
    \begin{equation}
        \left(\frac{1}{r}\partial_r r \partial_r +\frac{1}{r^2} \partial^2_{\theta} + \partial^2_z \right) \delta \phi = \frac{-q}{\epsilon_0} n_0(r) \int \mathrm{d}v ~\delta f
    \end{equation}
    
    Now, inserting $\delta f$ and  $\delta \phi$ leads to
    \begin{equation}
        \begin{split}
            \left(\frac{1}{r}\partial_r r \partial_r - \frac{m^2}{r^2} -k_n^2  \right) \frac{ \delta \phi_{nm}^{+}e^{ ik_n z } + \delta\phi_{nm}^{-}e^{-ik_n z }}{2}   \\
            = -\omega_{\mathrm{p}}^2 \int \mathrm{d}v \left[  \frac{ik_n\partial_v f_{\mathrm{v},0}}{2(p- im\omega_{\mathrm{D}} + ivk_n)}  \left(\delta\phi^{+}_{nm} + V(p)C^{+}_{nm} \right)e^{ ik_n z} \right. \\ \left. - \frac{ik_n \partial_v f_{\mathrm{v},0}}{2(p- im\omega_{\mathrm{D}} - ivk_n)}  \times \left(\delta\phi^{-}_{nm} + V(p)C^{-}_{nm} \right)e^{ -ik_nz} \right]
        \end{split}
    \end{equation}
    and using the fact that $\partial_v f_{\mathrm{v},0}$ is an odd function to:
    \begin{equation}
        \begin{split}
            \left(\frac{1}{r}\partial_r r \partial_r - \frac{m^2}{r^2} -k_n^2  \right) \delta\phi^{(p)}_{nm} \\
            = -\omega_{\mathrm{p}}^2 \int \mathrm{d}v  \frac{ik_n\partial_v f_{\mathrm{v},0}}{(p- im\omega_{\mathrm{D}} + ivk_n)}  \left(\delta \phi^{(p)}_{nm} + V(p)C_{nm} \right)  \\
             =  -K^2 \left( \delta\phi^{(p)}_{nm}(r) + V(p)C_{nm}\right)
        \end{split}
    \end{equation}
    
    For the radial profiles given in Eqs. \ref{equ: g(r)_inside} and \ref{equ: g(r)_outside} the l.h.s is given by:
    \begin{equation}
        \begin{split}
        \frac{1}{r}\partial_r~ r ~\partial_r ~\left( \delta\phi_{nml}^{(p)} g_{nml}(r)\right) \\= \left(-a_{nml}^2 + \frac{m^2}{r^2}\right)  \delta\phi_{nml}^{(p)} g_{nml}(r)
        \end{split}
    \end{equation}
    and the dispersion relation finally becomes:
    \begin{equation}
        \begin{split}
            k_n^2 D_{nml} \delta\phi_{nml}^{(p)}= K^2 V(p)C_{nml} \\
                \rightarrow \delta\phi^{(p)}_{nml} = V(p)C_{nml} \frac{K^2}{D_{nml}k_n^2}
        \end{split}
    \end{equation}
    with
    \begin{equation}
    D_{nml} = 1 + \frac{a_{nml}^2}{k_n^2} - \frac{\omega_{\mathrm{p}}^2}{k_n^2} \int \mathrm{d}v\frac{ik_n\partial_v f_{\mathrm{v},0}}{p- im\omega_{\mathrm{D}} + ivk_n} 
    \end{equation}
    and $K^2 = k_n^2 + a_{nml}^2$.
    The total eigenmode amplitude is a superposition of the amplitude by the distribution perturbation and the external field:
    \begin{equation}
        \begin{split}
            \delta \phi_{nml}(p) = \delta \phi^{(p)}_{nml} + V(p)C_{nml} \\
            = V(p)C_{nml} \left( 1+ \frac{K^2}{k_n^2 D_{nml}} \right)\\ =  V(p)C_{nml} \left(\frac{a^2_{nml} + k_n^2}{k_n^2} \frac{1}{ D_{nml}} \right)
        \end{split}
    \end{equation}
    Next, to facilitate the calculations, this equation is shifted in the $p$-coordinate from $p=i \omega_{nml} + \gamma_{nml} $ to $ p'=i \omega_{nml} + \gamma_{nml} - im\omega_{\mathrm{D}}$ and the above equation becomes:
    \begin{equation}
        \delta\phi_{nml}(p')  =  V(p')C_{nml} \left(\frac{a^2_{nml} + k^2}{k^2} \frac{1}{ D_{nml}(k_n,p')} \right)
    \end{equation}
    
    Using the convolution theorem, the temporal evolution can be calculated by:
    \begin{equation}
        \delta\phi_{nml}(t) = C_{nml} \int_0^t \mathrm{d}\tau V'(\tau) s_{nml}(t-\tau)
    \end{equation}
    
    where
    \begin{equation}
        V'(t)= \mathcal{L}^{-1}V(p') = \mathcal{L}^{-1}V(p) \mathrm{e}^{im\omega_{\mathrm{D}}t} = V(t) \mathrm{e}^{im\omega_{\mathrm{D}}t}
    \end{equation}
    
    and
    \begin{equation}
        s_{nml}(t) = \frac{a^2_{nml} + k_n^2}{k_n^2} \int^{\epsilon+i\infty}_{\epsilon-i\infty} \frac{\mathrm{d}p'}{2\pi i}e^{p't} \frac{1}{D_{nml}(k_n,p')} 
        \label{equ:g_nml(t)}
    \end{equation}
    
    The function $1/D$ is analytic everywhere except at its complex conjugated poles $p^{\pm}_{\mathrm{p}} = \pm i (\omega_{nml} - m\omega_{\mathrm{D}} ) + \gamma_{nml}$. 
    One can use Jordan's Lemma and the residue theorem to calculate the integral:
    \begin{equation}
        \int^{\epsilon+i\infty}_{\epsilon-i\infty} \frac{\mathrm{d}p'}{2\pi i}e^{p't} \frac{1}{D_{nml}(k_n,p')} = \sum_{p^{\pm}_{\mathrm{p}}} \mathrm{Res}(f,p_{\mathrm{p}})
    \end{equation}
    The first residuum is
    \begin{equation*}
        \begin{split}
            \mathrm{Res}(f,p_{\mathrm{p}}) = \mathrm{\lim_{p \rightarrow p_{\mathrm{p}}}} (p - p_{\mathrm{p}}) \frac{1}{D_{nml}(k_n,p)} e^{pt} \\
             = -\frac{(\omega_{nml}- m\omega_{\mathrm{D}})}{2i} \frac{k_n^2}{a^2_{nml} + k_n^2} e^{i(\omega_{nml}- m\omega_{\mathrm{D}} )t} e^{\gamma_{nml} t}
        \end{split}
    \end{equation*}
    using L'Hôpital's rule, the approximation $\partial_p D \approx -2 \omega_{\mathrm{p}}^2 / p^3$, low damping assumptions $p_{\mathrm{p}}^3 \approx - i (\omega_{nml}- m\omega_{\mathrm{D}})^3 $ and $\omega_{\mathrm{p}}^2 =(\omega_{nml}- m\omega_{\mathrm{D}})^2 (a^2_{nml} + k_n^2)/k_n^2$. Putting together both residues gives:
    \begin{equation}
        s_{nml}(t) =   - (\omega_{nml}- m\omega_{\mathrm{D}}) \times\sin{([\omega_{nml}- m\omega_{\mathrm{D}}] t)  } e^{\gamma_{nml} t}
    \end{equation}
    
    The temporal evolution of the mode amplitudes is then given by:
    \begin{equation}
        \begin{split}
            \delta \phi_{nml}(t) = -(\omega_{nml} - m\omega_{\mathrm{D}})  C_{nml} \\ \int_0^t ~\mathrm{d}\tau\left[  \phi_{\mathrm{rw}}(\tau) e^{-im(\frac{\omega_{\mathrm{rw}}}{m_{\mathrm{rw}}} - \omega_{\mathrm{D}} )\tau} \right.   \\ \left.
            \times \sin{((\omega_{nml} - m\omega_{\mathrm{D}})(t-\tau))} e^{\gamma_{nml} (t-\tau)}\right] ~~.
        \end{split}
    \end{equation}

\bibliography{literature}

\end{document}